# Seeing the Unheard: dynamics of thin liquid film in holographic ultrasonic field revealed by time-resolved Schlieren imaging

Weitao Sun[1,2]*, Diyao Wang[1], Yuheng Yang[1], Fangyu Cai[1], Mingchen Gao[1], Sirui Guo[1]

[1]School of Aerospace Engineering, Tsinghua University, Beijing, 100084, China

[2] Zhou Pei-Yuan Center for Applied Mathematics, Tsinghua University, Beijing, 100084, China

*Email: sunwt@tsinghua.edu.cn

## Abstract:

In this study, we introduce a unique approach that employs time-resolved Schlieren imaging to capture and visualize the dynamic changes of a thin liquid (mixture of water, soap and glycerin) film in ultrasonic wave field with high spatial and temporal resolution. By placing a soap film spanning a wire frame vertically in the path of light, we harnessed the vibrations induced by the ultrasonic waves, resulting in remarkable Schlieren imaging patterns. The investigation not only uncovers an unexpected branch flow phenomenon within the film, challenging existing assumptions, but also reveals a fascinating interplay between vortex flow and branch flow. The experiments have revealed a captivating spectrum of dynamic phenomena within the thin liquid films. The observation of small-scale capillary waves, large-scale standing waves, traveling waves, and the intricate fusion of capillary-gravity wave patterns underscores the rich complexity inherent in the interaction between the films and the holographic ultrasonic wave field. These diverse states of film dynamics provide a comprehensive understanding of the intricate interplay between various wave modes and fluid behavior, further enhancing comprehension of this fascinating phenomenon. The ability to visualize the pressure field opens up new avenues for optimizing acoustic levitation techniques, investigating particle behavior, and exploring potential applications in materials science and bioengineering.

## Introduction

In recent years, there has been growing interest in exploring innovative techniques to visualize and understand the acoustic pressure field[1,2]. Previous studies have primarily focused on indirect measurements and theoretical models to infer the pressure distribution[3,4]. However, directly visualizing the ultrasonic wave field has long posed a formidable challenge due to the inherent weakness of the ultrasonic pressure field, which induces only minute changes in the refractive index of air. These subtle variations are often insufficient to be captured by conventional imaging techniques, making it difficult to observe and analyze the intricate details of the ultrasonic wave field directly. As a result, innovative approaches and methodologies are required to overcome these limitations and enable the visualization and characterization of the elusive ultrasonic pressure field. Schlieren imaging is a powerful optical technique widely used in fluid dynamics and gas flow visualization[5,6]. It allows for the visualization of small changes in refractive index, providing insights into variations in density and temperature gradients within a medium. By employing a light source, typically a collimated beam, and an optical system with a knife-edge or a shearing plate,

Schlieren imaging can capture and amplify the subtle variations in refractive index, creating high-contrast images[7]. This technique has been extensively utilized to study shock waves[8], supersonic flows[9], combustion processes[10], and other phenomena involving density gradients. Time-resolved Schlieren imaging (TRSI) is a variant of the Schlieren imaging technique that enables the capture of dynamic events and temporal changes in refractive index[11]. By employing high-speed camera and synchronized illumination system, it allows for the visualization and analysis of fast-flowing processes, transient phenomena, and time-varying fluid structures. With its ability to reveal hidden flow structures and fine details, TRSI has become an indispensable tool for researchers and engineers in various fields of science and engineering.

Soap films and bubbles, composed of thin liquid films stabilized by surfactant molecules, possess remarkable properties that are deeply rooted in the principles of acoustics and optics[12-14]. With their high surface tension and low viscosity, these films serve as excellent mediums for visualizing and investigating intricate wave fields and complex flow phenomena[15]. By seeking to minimize interfacial energy, soap films inherently strive to reduce the surface area of their interfaces, leading to unique behaviors. This convergence of acoustics, optics, and chemistry in the study of soap films creates an interdisciplinary landscape, offering promising avenues for exploring and comprehending the intricate dynamics and phenomena within these films[16].

The study of vortex flow structures and vibrations in soap films has a rich history, with extensive theoretical and experimental investigations conducted by pioneers. Vibrating soap films have long been studied as archetypes of vibrating membranes, with early experiments conducted by Melde (1876) and observations reported by Taylor [17] and Bergman[18]. Over time, significant progress has been made in understanding the behavior of soap films subjected to vibration and flow. Lucassen et al.[19] developed a linear theory for the propagation of waves in soap films, while Afenchenko et al.[20] and Vega et al. [21] experimentally and theoretically explored the generation of strong recirculations in thin soap films. In contrast, Airiau[22], Rabaud [23] and Brazovskaia et al. [24] observed and investigated the self-adaptation of thick soap films to forcing, showcasing their ability to sustain large amplitude oscillations at different frequencies.

While significant progress has been made in understanding the quasi-steady flow and vibration phenomena in soap films, the dynamics resulting from the interaction between soap films and the holographic ultrasonic field is largely unexplored. A complete model for the vibration-convection phenomena in soap films has remained elusive [25]. The coupled vibration and flow modes of soap films under the influence of the holographic ultrasonic field of an acoustic levitator have not been studied, resulting in a limited understanding of the phenomena in such experiments. This knowledge gap arises due to the inherent complexities associated with this coupled vibration-convection phenomenon [26].

In this study, we embark on investigating the previously unexplored dynamics of soap films in the holographic ultrasonic field. The aim is to bridge the gap in knowledge by combining TRSI with soap film, which serves as a sensitive indicator of pressure variations. By leveraging the well-established technique of Schlieren imaging, we can visualize density variations in transparent media and, in this case, capture the intricate pressure patterns within the levitation wave field with high spatial and temporal resolution.

Through a series of experiments, we placed soap film in the ultrasonic wave field of an acoustic levitation apparatus and recorded the dynamic changes using a high-speed camera. With the help of the remarkable sensitivity of soap films to pressure variations, TRSI emerges as a powerful tool for capturing density variations in transparent media, enabling us to visualize the previously unseen details of the flow and vibration of soap film caused by ultrasonic wave field. It is within this unseen realm that we encounter a surprising and remarkable phenomenon—an intricate interaction and coupling between the capillary wave and gravity wave. Various intriguing observations have been made concerning the diverse states exhibited by the soap film as it moves in the acoustic levitation wave field. These observations encompass, but are not limited to, phenomena such as small-scale capillary waves, large-scale standing waves, traveling waves, and intricate combinations of capillary-gravity wave patterns.

While this work represents a significant step forward in visualizing the acoustic levitation pressure field, there are certain limitations to be acknowledged. The use of soap film as an indicator of pressure variations introduces some complexities and potential artifacts. Additionally, the experiments were conducted under specific conditions and may not capture the full range of pressure patterns that can occur in different levitation setups.

In future research, we intend to further refine and validate our visualization technique, explore the influence of different parameters on the pressure distribution, and investigate the potential applications of the pressure field visualization in enhancing levitation capabilities and manipulating particles with greater precision.

## Method

The experimental setup consisted of a TRSI system and an acoustic levitation system. The TRSI system comprised an LED light source, two convex lenses, and a high-speed camera (Figure 1). The lenses were carefully aligned to create a parallel light beam, which was directed through the acoustic levitation system and onto the soap film. The LED light source emitted white light and had a power output of 15 W. The camera used in our setup was the Chronos 2.1, offering frame rate ranging from 1000 to 24046 frames per second (FPS). The high-speed camera was synchronized with the LED light source to capture the dynamic changes in the soap film with high temporal resolution.

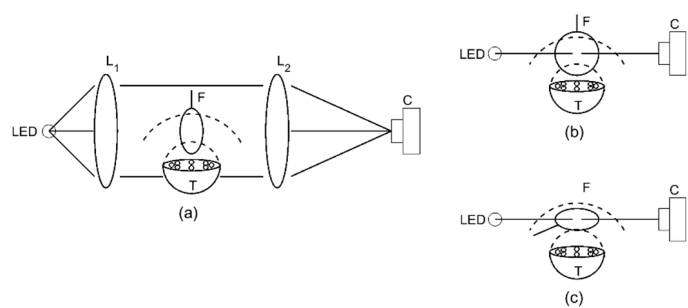

Figure 1. Schematic diagram of the experiment setup. A white LED point light source with a power of 15W is used. The soap film (F), positioned vertically, is placed between two lenses $L_1$ and $L_2$. The first lens collimates the light, while the second lens focuses it. The ultrasonic transducers (T) generates disturbances and variations of the refractive index in the air around the soap film. The deviated light forms an image at the focal point of the second lens, which is captured by a high-speed camera (C) for subsequent analysis.

The holographic ultrasonic field is generated through the employment of an acoustic levitation apparatus[27]. The acoustic levitation system was constructed using 30 ultrasonic transducers (16 mm, 40 kHz). These transducers were arranged in a bowl-shaped container, fabricated using a 3D printer, to generate a uniform levitation wave field (Figure 2). The system was powered by a 7.4V lithium battery and controlled using an Arduino Nano microcontroller. The L298N dual motor driver board was utilized for transducer control and power regulation.

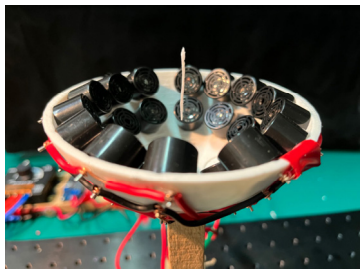

Figure 2. Acoustic levitation of polystyrene rod.

To create the soap film, a circular frame was employed. A soap solution was prepared by mixing soap concentrate, distilled water, and glycerin. The frame was dipped into the soap solution, allowing the soap film to form. Care was taken to ensure a thin and uniform film by gently blowing air across the frame. The soap film was carefully placed in the center of the transducer array, perpendicular to the direction of the light beam (Figure 1). The experimental setup was calibrated to optimize alignment, focus, and exposure settings of the camera for optimal image capture. The acoustic levitation system was activated, generating the levitation wave field and inducing pressure variations in the air around the soap film (video 1). The high-speed camera recorded the dynamic changes in the soap film as it interacted with the pressure variations in the levitation wave field.
Multiple experiments were conducted, adjusting soap film positions to capture a range of pressure distribution patterns and fluid flow dynamics. The recorded images were subjected to appropriate image processing techniques to visualize and analyze the pressure distribution within the levitation wave field. Quantitative measurements of pressure gradients and fluid flow velocities were performed using suitable analysis tools, contributing to a comprehensive understanding of the system dynamics.

**Results**

When the acoustic levitation system was switched off, the soap film displayed no distinct patterns. However, upon powering on the system, a sudden appearance of a dark region in the film indicated the presence of high-pressure areas within the levitation wave field (Figure 3). This observation confirmed the sensitivity of the soap film as an indicator of pressure variations. Furthermore, the pressure field generated by the acoustic levitation system induced complex fluid flows and convections within the soap film. The flow field exhibited a rich variety of convection patterns throughout the entire film, indicating the intricate dynamics of the system (Figure 3).
The captured Schlieren images revealed conspicuous dark regions within the soap film, offering a visual representation of localized high-pressure zones within the acoustic levitation wave field. In addition, upon examining the time-resolved Schlieren snapshots, it becomes evident that these images unmistakably depict the emergence of vortex traps within the dynamic environment of the holographic ultrasonic wave field. Within the central region of the soap film, a discernible dark area becomes apparent, marking the cross-sectional view of the rod-shaped trap responsible for securely holding the polystyrene particles in place. It is worth noting that the acoustic radiation force effectively converges towards the central vertical axis contained within this dark rod-shaped structure.

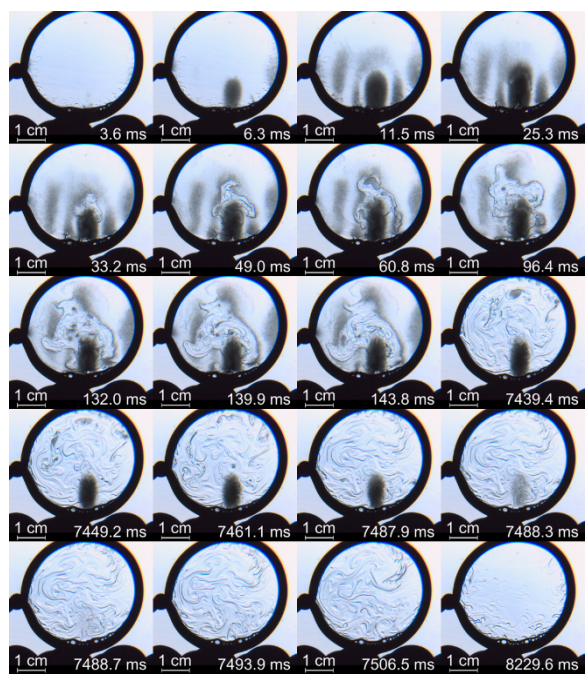

Figure 3. Snapshots of Schlieren images of a liquid film spanning a wire frame in the ultrasonic wave field of the acoustic levitation system. The images reveal the dynamic flow structures, convection phenomena, and localized dark regions representing high acoustic pressure areas.

Surrounding the central dark region, numerous minor dark areas emerge, providing clear evidence of the intricate and multifaceted nature of the holographic wave field. Notably, the closest minor region even assembles to form a closed concentric layer positioned just outside the boundaries of the central trap. This phenomenon underscores the complexity and richness of the acoustic levitation pressure patterns engendered by the holographic ultrasonic wave field.

It is evident that the central trap initiates fluid motion within the film. This flow gradually extends outwards, giving rise to intricate vortices that span the entire soap film's surface. The presence of these flow patterns and convections in the soap film offers valuable insights into the behavior of the pressure field in acoustic levitation. These observations highlight the complexity and dynamic nature of the levitation process, contributing to a deeper understanding of the phenomenon.

The experimental results revealed significant insights into the visualization of the acoustic levitation pressure field using time-resolved Schlieren imaging of a soap film. The Schlieren images captured the dynamic changes in the soap film as it interacted with the pressure variations within the levitation wave field. The results effectively reveal contrasting dark and bright patterns, a result of thickness variations and ripples present on the thin film's surface. Yet, capturing intricate details of dendritic patterns within the images poses challenges due to their rapid fluctuations. Moreover, these patterns are confined to the two-dimensional film plane. Conventional Schlieren imaging lacks the capability to observe height variations perpendicular to the film surface, limiting the comprehensive observation of the system's behavior.

We conducted an in-depth examination of minute surface height variations by employing high-speed photography. The camera's view direction was maintained nearly parallel to the film surface. To illuminate the film surface effectively, a high-intensity photography light was utilized. The light reflected off the film surface was directed into the camera, allowing us to capture intricate surface motions with precision and detail. Figure 4illustrates the dynamic behavior of the film surface when subjected to stimulation by the holographic ultrasonic wave.

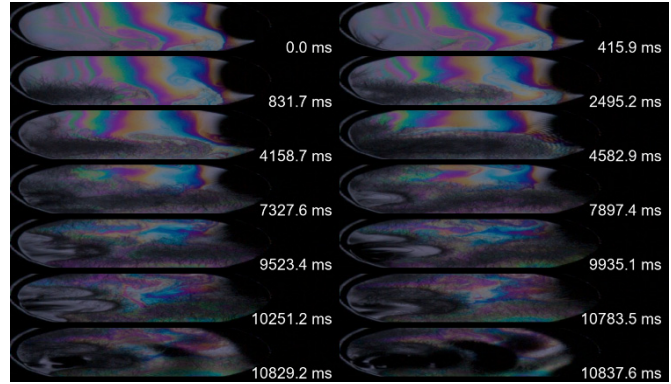

Figure 4. Snapshots of film motions in the ultrasonic wave field. The images are captured by high speed camera with a frame rate of 24046 fps.

The findings from this study not only expand our knowledge of acoustic levitation but also have broader implications for fields such as particle manipulation, microfluidics, and acoustic wave research. The ability to visualize and characterize the pressure field in a non-invasive manner opens up possibilities for advanced control and optimization of acoustic levitation systems.

In Figure 5, we observed the presence of a traveling wave on the thin film's surface. Over approximately 4.86 milliseconds, the film's shape underwent a complete cyclic transformation. This motion involved the movement of wave peaks from the center towards the frame's edges and then reflecting back to the central portion of the film after reaching the edges.

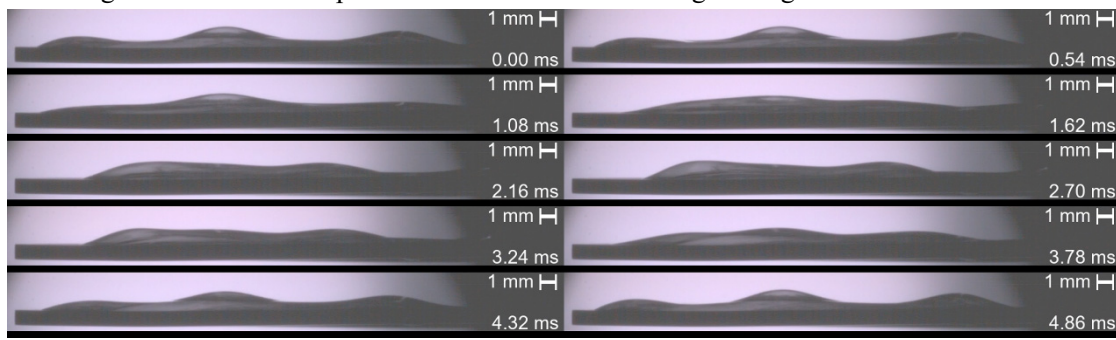

Figure 5. Snapshots of traveling wave on the film. The traveling wave exhibits a cyclic motion with a period of approximately 4.86 milliseconds. The images are captured by high speed camera with a frame rate of 14825 fps.

We also observed traveling waves with different period. During more intense vibrations of the thin film, the traveling waves exhibited larger amplitudes and longer periods. Figure 6illustrates that the traveling wave has a period of approximately 13.76 milliseconds. Additionally, the film surface experiences such vigorous oscillations that small droplets are expelled from its surface.

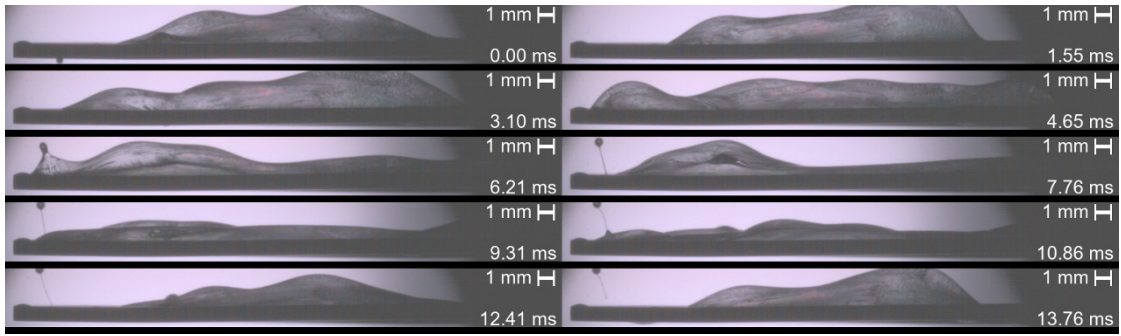

Figure 6. Snapshots of traveling wave on the film. The traveling wave exhibits a cyclic motion with a period of approximately 13.76 milliseconds. The images are captured by high speed camera with a frame rate of 14825 fps.

We also observed standing waves on the film when it was subjected to the holographic wave field. Figure 7 depicts the surface with multiple peaks. Remarkably, these peaks remained stable for an extended period, approximately 667.79 milliseconds. Such long-term waveforms represent intrinsic vibrational modes of the film itself when it is located at a specific position in the holographic wave field.

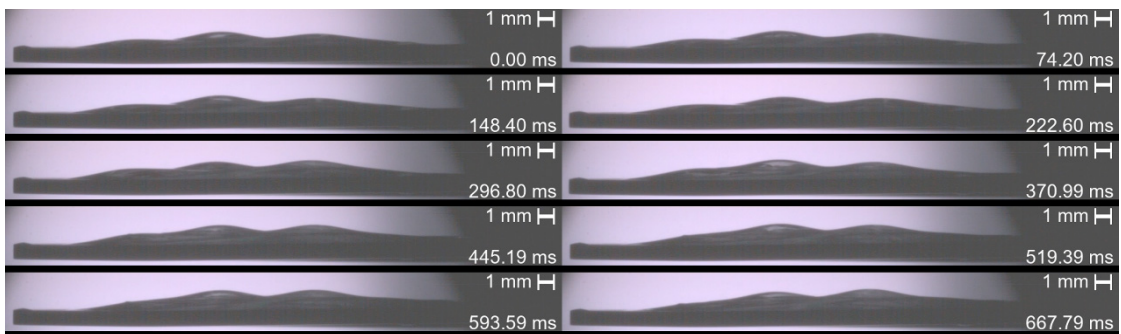

Figure 7. Snapshots of standing wave on the film. The standing wave exhibits a period of approximately 667.79 milliseconds. The images are captured by high speed camera with a frame rate of 14825 fps.

In the observations, we also identified the presence of capillary-gravity wave patterns on the film. This intriguing phenomenon is characterized by the coexistence of large-scale gravity waves forming the background landscape and smaller capillary waves propagating atop the surface of the gravity waves (Figure 8). The large-scale gravity waves exhibit a wavelength approximately matching that of the ultrasonic wave, measuring around 8.6 mm. These waves, while grand in scale, propagate at a relatively slower pace. In contrast, the small ripples created by the capillary waves are minuscule when compared to the grandeur of the gravity wave. Remarkably, these small-scale capillary wave ripples propagate at a considerably faster pace than their larger counterparts in the gravity wave.

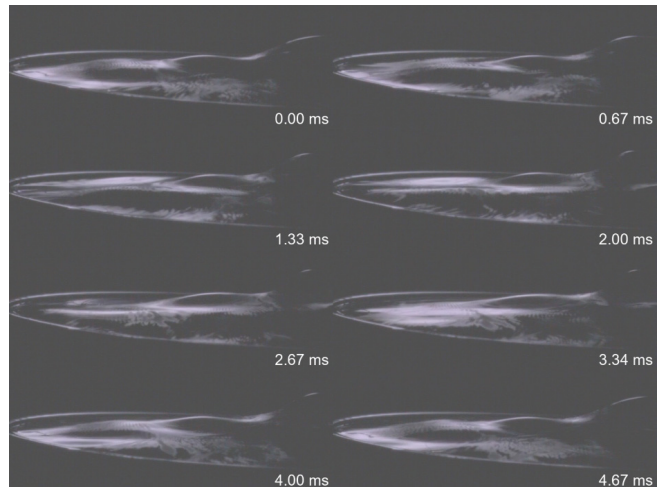

Figure 8. Snapshots of capillary-gravity wave patterns on the film. The capillary-gravity wave pattern consists of large-scale gravity waves forming a backdrop with smaller, fast-moving capillary waves rippling across their surface. The images are captured by high speed camera with a frame rate of 10488 fps.

**Discussion**

The discovery of rich and fascinating dynamic flow patterns in this study represents a significant contribution to the understanding of acoustic levitation phenomena. Through the interplay of ultrasonic pressure vibrations, light interferences, and the unique mechanical properties of the soap film, we observed the emergence of intricate capillary waves and oscillatory motions within the film. This intricate interaction between the ultrasonic wave and the liquid membrane gave rise to the formation of vortices and fascinating flow patterns in regions undergoing vibration. These findings not only provide valuable insights into the complex dynamics of acoustic levitation but also highlight the importance of considering the interplay between mechanical forces and liquid behavior in such systems.

Surprisingly, the soap film stimulated by holographic ultrasonic wave exhibits fascinating patterns of dendritic formation. The formation of dendritic patterns arises from capillary waves, which induce delicate ripples on the surface of the thin liquid film. These ripples exhibit variations in thickness, leading to the complex dark/bright fringes when the liquid film is visualized through TRSI. These remarkable fractal-like patterns bear a striking resemblance to the intricate branching patterns seen in branch flow phenomena. However, the precise connection between these patterns and the phenomenon of branch flow remains enigmatic, with much of the underlying relationship yet to be comprehensively understood.

The dendritic patterns manifest not only in regions characterized by disrupted vibrations but also emerge at the very core of the flowing vortex. Such patterns persist as they journey alongside the flowing vortex through space, gradually dissipating over time. This mesmerizing interplay between vortex flow and dendritic patterns reveals a new layer of complexity within the dynamics of soap film in the ultrasonic wave field. Understanding the intricate interplay between waves and delicate films holds great promise in various scientific and engineering disciplines.

Furthermore, the complex fluid flows and convections observed in the soap film highlight the dynamic nature of the acoustic levitation system. This phenomenon has significant implications for the manipulation and control of particles in various applications, including microfluidics, particle sorting, and acoustic trapping. These findings contribute to the knowledge base by revealing the intricate details of fluid flow dynamics within the levitation wave field, enabling researchers to optimize levitation conditions and enhance particle manipulation capabilities.

Possible explanations for unexpected or interesting observations, such as the formation of unique flow patterns, could be attributed to the interplay between pressure variations and fluid properties within the soap film. Further analysis and modeling efforts can shed light on these phenomena and provide a deeper understanding of the underlying mechanisms.

Comparing these results with previous studies, it is evident that the application of time-resolved Schlieren imaging to visualize the acoustic levitation pressure field is a novel and promising approach. While previous works have focused on visualizing strong wave fields generated by large transducers, this study extends the application of Schlieren imaging to the relatively weaker pressure field of acoustic levitation. This demonstrates the versatility and effectiveness of the technique in capturing subtle pressure variations.

This study successfully visualizes the acoustic levitation pressure field through TRSI of a soap film. By employing TRSI, we have provided a unique and comprehensive visualization of the intricate dynamics and flow phenomena within the liquid film under the influence of the acoustic levitation pressure field. This discovery not only enhances our understanding of the fundamental principles governing acoustic levitation but also paves the way for further investigations and advancements in this field. The insights gained from this study have the potential to drive future research directions and inspire innovative applications in various scientific and technological domains.

## Conclusion

In this study, we have delved into the intricate dynamics of thin liquid films within holographic ultrasonic wave fields, employing the technique of time-resolved Schlieren imaging. The findings have shed light on the previously unexplored interaction between these films and the complex pressure variations induced by the holographic ultrasonic waves. Through a series of experiments, we have successfully visualized fluid flow dynamics on soap film driven by the ultrasonic wave field. The utilization of time-resolved Schlieren imaging has proven to be instrumental in capturing the subtle film surface variations and revealing dynamic flow patterns. The observed dark regions within the soap film have confirmed its sensitivity as an indicator of pressure changes. Moreover, the intricate flow structures and convection patterns observed in the soap film provide crucial insights into the behavior of the acoustic levitation wave field. An unexpected yet captivating discovery emerged in the form of dendritic patterns that manifest not only in regions of disrupted vibrations but also at the heart of flowing vortices. This phenomenon, along with the interplay between capillary and gravity waves, has opened a new realm of complexity within the dynamics of soap film vibrations. While our study contributes significantly to the understanding of acoustic levitation phenomena and wave interactions, further research is essential to comprehensively unravel the underlying mechanisms. This work holds implications not only for fundamental research but also for practical applications in various fields such as microfluidics, particle manipulation, and materials science. The ability to visualize and manipulate the intricate pressure and flow dynamics in ultrasonic wave fields provides opportunities for enhanced control and optimization in various engineering and scientific domains.